\documentclass[a4paper,11pt]{article}
\usepackage{amsmath}
\usepackage{amssymb}
\usepackage[normalem]{ulem}
\usepackage{graphicx}
\usepackage{placeins}
\usepackage{multicol}
\usepackage[margin=2cm]{geometry}
\usepackage{enumitem}

\begin{document}

\title{The Undulator Radiation Collider: An Energy Efficient Design For A $\sqrt{s}=10^{15}\, \mathrm{GeV}$ Collider}

\author{Francis Bursa\\
School of Physics and Astronomy,
University of Glasgow,\\
Glasgow G12 8QQ, United Kingdom \\
\\
francis@semichrome.net}
\maketitle

\begin{abstract}
We discuss the main factors affecting the design of accelerators aiming to investigate physics at the GUT scale. The most important constraints turn out to be the energy used and the time taken to accumulate sufficient luminosity. We propose a photon collider design, where the photons are generated by undulator radiation from high energy muon beams. This reduces the energy requirements by a factor of more than $10^7$ compared to a \emph{pp} collider. Much of the reduction in energy use is achieved by using a periodic magnetic field, and by splitting the muon wavefunctions spatially to reduce the photon beam divergence; these prevent a cascade of secondary reactions at the collision points The proposed collider would be powered by (part of) a Dyson swarm constructed around the Sun, and efficient use of energy will be important to reduce the time needed to reach the desired number of collisions. We also discuss why a neutrino collider would be much less efficient.
\end{abstract}

\begin{multicols}{2}

\section{Introduction}

In the distant future we may wish to construct a collider with a centre of mass energy $\sqrt{s}$ around the scale of grand unified theories (the GUT scale; approximately $10^{15}$\,GeV) in order to investigate physics at these energies directly. This paper discusses the problems affecting the design of such a collider and proposes an undulator radiation collider as an energy efficient solution.

Throughout this paper, we will assume the Standard Model is correct up to the GUT scale; otherwise the design of (and motivation for) a GUT scale collider would depend on the new physics. 
The remainder of this section justifies the energy and luminosity required for a GUT scale collider. Section~\ref{particlechoice} explains why a photon collider is a good choice, and describes a method for achieving a photon beam with narrow bandwidth and low divergence. Section~\ref{interaction region} describes a potentially serious problem due to secondary reactions in the interaction region and how this can be mitigated by using a periodic magnetic field in the interaction region. This section also explains why the same approach cannot be used for neutrino colliders.

The detectors and accelerators are described in Sections~\ref{detector} and~\ref{accelerator} respectively, and Section~\ref{energy} describes the Dyson swarm and energy transmission. Finally we sum up in Section~\ref{conclusions}.

\subsection{Target energy and luminosity}
\label{targetparams}
\subsubsection*{Energy}

After the discovery of the Higgs boson, it is not clear at what energy scale new physics might next be found. Naturalness suggests that this scale should be as low as possible, and preferably at the weak scale ($\sim100$\,GeV); however this is already in tension with current limits and it is possible that naturalness arguments are simply wrong. Both dark matter and dark energy require physics beyond the Standard Model, but this physics might occur at any scale.

An indication for a scale comes from the fact that the three Standard Model couplings become approximately equal in the range $10^{14}$--$10^{15}$\,GeV. This observation led to the idea of grand unified theories, in which the three Standard Model gauge groups become unified into one around this scale. The simplest non-supersymmetric GUTs have been ruled out, and in this article we are assuming low-energy supersymmetry does not exist. However, more complicated non-supersymmetric GUTs are still possible~\cite{Agashe:2014kda}, and in any case the crossing of the couplings does suggest that something interesting may happen at this scale. 

Neutrino masses also provide an indication of a scale at which new physics might be expected. If the see-saw mechanism is correct, these would be generated at a scale of around $10^{14}$\,GeV~\cite{Agashe:2014kda}. The see-saw mechanism can only work if neutrinos are Majorana particles, which is currently unknown; however, it is another indication that there may be new physics in the range $10^{14}$--$10^{15}$\,GeV.

To be specific, in this paper we will adopt a target energy of $10^{15}$\,GeV, at the upper end of the ranges above. However, the design proposed here could easily be adapted to somewhat lower or higher energies.

\subsubsection*{Luminosity}
In addition to having sufficient energy, a collider must have a large enough luminosity to discover new physics. The required integrated luminosity depends on the cross-section of interest, which is naturally of order
\begin{equation}
\sigma \sim \frac{\alpha^2}{E^2},
\end{equation}
where $\alpha$ is the relevant coupling constant and we assume that the leading order Feynman diagram includes two vertices. Taking, say $\alpha = 0.1$, we find $\sigma = 4\times10^{-64}\,\mathrm{m}^2 = 4\times10^{-36}$\,barn.

The number of events observed is $N=\sigma L$, where $L$ is the integrated luminosity. Assuming we want to observe at least 10 or so events, this translates to the requirement $L \gtrsim 2.5\times10^{64}\, \mathrm{m}^{-2} = 2.5\times10^{18}\,\mathrm{ab}^{-1}$. To put this into perspective, the total integrated luminosity at the LHC is expected to be around $3\,\mathrm{ab}^{-1}$.

\section{Choice of particles}
\label{particlechoice}
We now turn to the choice of particles to collide. All colliders to date have used either electrons or protons (and sometimes their antiparticles). Of these protons are clearly better at high energy since the synchrotron radiation losses are smaller by a factor $\left( \frac{m_p}{m_e} \right)^4 \approx 10^{13}$. This is a significant consideration even for a linear collider, since the beams have to be bent in order to be focused at the interaction points. For example, for electrons with energy $5\times 10^{14}$\,GeV, a beam of width $1\,\mathrm{\mu m}$ would have to be focused over a length of at least $10^{11}$\,m to avoid losing most of its energy by synchrotron radiation.

The total cross-section at $\sqrt{s} = 10^{15}$\,GeV for proton-proton collisions is approximately 1200\,mb~\cite{Agashe:2014kda}. For an integrated luminosity of $2.5\times10^{18}\,\mathrm{ab}^{-1}$, this corresponds to a total of $3\times10^{36}$ collisions over the lifetime of the experiment. The energy required per collision is $10^{15}$\,GeV (assuming 100\% efficiency!); therefore the total energy required is approximately $5\times10^{41}$\,J. Since the luminosity of the Sun is $3.8 \times10^{26}$\,W, the total energy output of the Sun would be required for some 40 million years.

This is a lot of energy, and it would be nice to be able to use a bit less. The energy required is proportional to the total cross-section, so we should aim to use particles with as small a cross-section as possible.

Using electrons is almost certainly impossible due to the huge synchrotron radiation losses. For other charged particles $X$, the process $X^+ X^- \to X^+ X^- e^+ e^-$ will contribute to the total cross-section. To leading order, the cross-section for this process is~\cite{Budnev:1974de}
\begin{equation}
\sigma_{X^+ X^- \to X^+ X^- e^+ e^-} = \frac{28 Z^4 \alpha^4}{27 \pi m_e^2} \left( \mathrm{ln} \left(\frac{2E}{m^2}\right) \right) ^3,
\label{Charged cross-section}
\end{equation}
where $E$ is the beam energy and $Z$ is the charge of $X$. This decreases only very slowly with the particle mass $m$. For muons it is 500 mb and for $\tau$ particles it is still 400 mb, so the gain compared to protons is relatively small. Many hadrons are heavier, but these will have large additional contributions to the cross-section from the strong interaction. The $W$ boson is heavier still, but its lifetime is far too short. So no known charged particle will do.

The alternative is to collide neutral particles, which can have far smaller cross-sections. However, they cannot be accelerated electromagnetically. Instead they must be produced indirectly, or produced at low energy and then accelerated by colliding them with a high energy beam.

Neutrinos have an extremely low cross-section (220 pb for $\nu \bar{\nu}$ and 70 pb for $\nu \nu$~\cite{Lacki:2015lda}), and high energy neutrinos can be produced by the decay of high energy muons. However, there is a problem at the energies we are considering -- muons are too stable! At an energy of $5\times 10^{14}$\,GeV, muons travel approximately $3 \times 10^{18}$\,m (300 light years) before they decay, requiring a collider at least twice that size (one decay length for each arm). Even worse, the divergence (angular spread) of the neutrino beam, which is of order $10^{-16}$, will result in very wide neutrino beams at the interaction points. 

There is in principle an alternative method of producing high-energy neutrinos: we could collide particle-antiparticle (say $\mu^+\mu^-$) pairs on almost parallel trajectories such that their energy in the centre of mass frame is equal to the $Z^0$ mass. They will produce $Z^0$ bosons with a fairly large cross-section. These will subsequently decay to a mixture of neutrinos, leptons and quarks, and their antiparticles. The leptons and quarks can be filtered out by passing the beam through a long (tens of kilometers) absorber, leaving a nearly pure beam of neutrinos and antineutrinos. This will cost a factor of a few in efficiency, but it might be worth it since the neutrino cross-section is so small.

To obtain (anti-)neutrinos with an energy of $5\times 10^{14}$\,GeV the initial colliding particles must have the same energy. Their transverse momentum must be near $\frac{m_{Z^0}}{2} \approx 46$\,GeV, and the neutrinos and antineutrinos produced will have transverse momenta of the same magnitude, so in this case the divergence of the beam will be larger, roughly $10^{-13}$. This is not a problem in terms of the width of the beams, which would still only be nanometres wide after an absorber tens of kilometres long. However, the divergence nevertheless does cause problems, since it means that secondary particles produced at small angles in the collisions remain in the beam and will interact with the (anti-)neutrinos, leading to a large increase in the effective cross-section. We will discuss this in more detail in Section~\ref{interaction region} below, where we will show how this increase can be avoided for photons, but unfortunately not for neutrinos.

Further alternatives we might consider are, for example, the reaction $\mu^+ \mu^- \to H^0$ to produce high energy Higgs bosons some centimetres from the collision point (Higgs bosons have a range of approximately 20 cm at $5\times 10^{14}$\,GeV) or $\mu^+ \mu^- \to \Upsilon(3S)$ to produce high energy bottomonium tens to hundreds of metres from the collision point ($\Upsilon(3S)$ particles have a range of approximately 500 m at $5\times 10^{14}$\,GeV). Higgs bosons are weakly-interacting, and bottomonium, although strongly interacting, is spatially much smaller than a proton, so both these options would reduce the total cross-section. With this method the particles should be produced at resonance so that energy is not wasted producing other, unwanted, types of particles. Nevertheless, for Higgs bosons the process will be very inefficient: the cross-section for $\mu^+ \mu^- \to H^0$ at resonance is 40 pb~\cite{Neuffer:2015nna}, whereas the cross-section for process~(\ref{Charged cross-section}) is 2~mb at the same centre-of-mass energy, giving a Higgs boson production efficiency of 0.000002\%. Similarly, $\Upsilon(3S)$ production would also have a low efficiency.
 
Another problem is that there may be interactions between the produced particles and the incoming beams, e.g. the reaction $H^0 \mu \to \mu W^+ W^-$, which would reduce the intensity of the beam produced. Whether such reactions are a serious problem requires further investigation.

\subsubsection*{Photons}

The cross-section for $\gamma \gamma \to \mathrm{hadrons}$ at $\sqrt{s} = 10^{15}$\,GeV is only about 0.012 mb~\cite{Agashe:2014kda}. The reaction $\gamma \gamma \to e^+ e^- e^+ e^-$ contributes an additional 0.006 mb~\cite{Ginzburg:1981vm}, giving a total of 0.018 mb. This reduces the energy requirements for a photon-photon collider by a factor of 70000 compared to protons and 30000 compared to muons.

$\gamma \gamma$ colliders have been proposed with centre of mass energies in the GeV to TeV range (see e.g.~\cite{Asner:2001vh, Bogacz:2012fs}). The high energy photons in these designs would be produced by inverse Compton scattering: a low energy laser beam would be bounced off a high energy electron beam to produce high energy photons. The limit on this process is pair production of electron-positron pairs by interactions between the incoming and outgoing photons; to avoid this, one must have
\begin{equation}
x = 4 \frac{E_{e^-} E_{\gamma, in}}{m_e^2} < 4.8.
\end{equation}
The maximum outgoing photon energy is given by $E_{\gamma, out} = \frac{x}{x+1} E_{e^-}$, and is 83\% of the electron beam energy for $x=4.8$.

To use this method at extremely high energies, it seems all we have to do is to bounce the photons off muon beams rather than electron beams to avoid the issue of synchrotron losses. Unfortunately, this doesn't work. The limit $x=4.8$ applies when the masses of the particles being pair-produced and those in the incoming beam are the same. This is the case for photon collider designs using electrons, but not for our situation: the incoming beam contains muons but the limit on $x$ is given by the lightest charged particles that can be pair-produced, which are electrons. The limiting value of $x$ in this situation is
\begin{equation}
x=2 \left( \frac{m_e}{m_\mu}\right) ^2 + 2 \sqrt{\left( \frac{m_e}{m_\mu}\right) ^2 + \left( \frac{m_e}{m_\mu}\right) ^4} \approx 0.0097.
\end{equation}
Therefore the maximum outgoing photon energy is over 100 times smaller than the muon beam energy. This means an increase in energy requirements by a factor of 100, reducing the gain to a factor of 700 compared with protons, or 300 compared with muons. Furthermore, synchrotron losses will increase by a factor of $10^8$, unless the curvature radii are increased by a factor of 10000 from the already very large radii required at a beam energy of $5\times 10^{14}$\,GeV, and the accelerating sections will need to be 100 times longer. These disadvantages seem likely to outweigh the benefit of a reduced energy requirement.

Fortunately, there is an alternative mechanism for producing high energy photons: synchrotron radiation. The question is whether two sufficiently intense beams of synchrotron radiation can be collided in a small enough area. To achieve this it is desirable for the photons to be emitted into a narrow cone and for the emission region to be short so that the beam does not diverge very much. The angular width most of the synchrotron radiation is emitted into is $\theta \sim 1/\gamma$, which is indeed very small for the energies we are concerned with. To reduce the length of the emission region the magnetic field strength should be large.

For the analysis of the collisions it would also be desirable for the photon spectrum to be narrow, and to reduce the length of the accelerator the photon energy should not be too far below the energy of the charged particle. Both of these properties can also be achieved by using strong magnetic fields. Specifically, the magnetic field strength in the rest frame of the radiating particle, in units of the critical field strength, is
\begin{equation}
\chi = \frac{\gamma B}{B_0}
\end{equation}
where $B_0$ is the QED critical magnetic field strength, which is $4.4 \times 10^9$\,T for electrons and scales as $m^2$.  For $\chi < 1$, the synchrotron radiation has a broad spectrum centred around $\chi E$, (where $E$ is the energy of the charged particle), but for $\chi \ge 1$, the spectrum becomes increasingly narrow and peaks just below $E$ -- see, for example, Figure 15.1 in~\cite{Uggerhoj}. Therefore passing a beam of charged particles through a very strong magnetic field would produce the desired photon spectrum.

However, there is a serious difficulty with this. If the magnetic field is stronger than the critical field strength for the radiating particles, which must be muons or heavier particles (electrons presumably being ruled out as discussed above), it will be very much above the critical field strength for electrons. Therefore the photons will be able to pair-produce electron-positron pairs. The photon beam will be attenuated, with the attenuation coefficient per unit length given by~\cite{Homola:2013sya}
\begin{equation}
\alpha = \frac{\alpha_{EM}m_ec}{2 \hbar}\frac{B}{B_0}T(\chi),
\end{equation}
where $T(\chi) \approx 0.60 \chi^{-1/3}$ for large $\chi$, and here $\chi = \frac{0.5 E_\gamma}{m_e c^2}\frac{B}{B_0}$ and the critical field strength for electrons should be used. The length scale over which a particle of energy around $5 \times 10^{14}$\,GeV loses most of its energy to synchrotron radiation is rather small even for moderately strong magnetic fields, $l_\mathrm{loss} \approx \frac{400}{B^{2/3}}$\,m, where $B$ is measured in Teslas. Over this length scale the loss due to electron-positron pair production is significant: the photon intensity is attenuated to $e^{-l_{loss} \alpha}\approx 30\%$. Some of the electrons and positrons will then radiate again, so the overall efficiency is slightly better than this, but this is still rather inefficient.

\subsubsection*{Undulator design}

An alternative way to achieve a narrow spectrum is to use an undulator -- a setup in which the magnetic field is periodic with period $\lambda_u$ and causes the charged particles to have a periodic trajectory, usually sinuisoidal. If there are $N$ periods the bandwidth of the emitted radiation (on axis) is $\sim \frac{1}{N}$ centred at a wavelength approximately given by
\begin{equation}
\lambda \approx \frac{\lambda_u}{2 \gamma^2},
\label{Undulator equation}
\end{equation}
where $\gamma$ is the Lorentz factor of the primary particles. Note that the radiated wavelength is independent of the magnetic field strength; however, the radiated intensity will be very small unless $B$ is such that the corresponding synchrotron radiation spectral density is appreciable at $\lambda$. Thus if we do not want $E$ to be very much greater than $E_\gamma$, we must have $\chi$ not much smaller than 1. 

The best particles to use as the primaries are probably muons. This is because (again, assuming electrons are impractical) all other particles that are sufficiently stable to accelerate are hadrons, for which pion emission~\cite{Maruyama:2015bga} would compete with synchrotron radiation.
 
Ideally, the muon energy $E$ would be only slightly higher than $E_\gamma$. In this case recoil effects will be very significant. These will reduce the coherence of the undulator~\cite{Bulyak:2015yua}: after the muon has emitted a photon its velocity will be reduced and the relative phase between the muon and photon will no longer increase by $2 \pi$ over one undulator period. However, this does not matter: most of the initial muon energy will already have been transferred to the first photon emitted, so what happens to the muon after that is not important (and much of its remaining energy can be recovered in the absorbers described in section~\ref{detector}).

A further advantage of an undulator is that it will suppress pair production. This is because the relative phase between the photon and the electron--positron pair will in general not be a multiple of $2 \pi$ per undulator period, so over many periods there will be destructive interference, reducing the amplitude for pair production.

The undulator parameters --- the muon energy, magnetic field, number of periods, and shape, which need not be sinusoidal --- will determine the photon energy, photon bandwidth, angular width of the photon beam, and the pair production rate. The actual choice of parameters will be a trade-off between these properties and will depend on the details of future technological development. However, a guess at reasonable parameters shows that they do not seem to be too challenging; indeed, some of them are already possible with present-day technology. For example, for a muon energy of $6 \times 10^{14}$\,GeV, the undulator wavelength, given by~(\ref{Undulator equation}), is 160 m, and the required magnetic field is 0.02 T. The energy loss distance with this field is $\approx 7000$\,m, which would give $N \approx 50$ undulator periods. 

\begin{figure}[h]
\centerline{\includegraphics[width=0.6\linewidth]{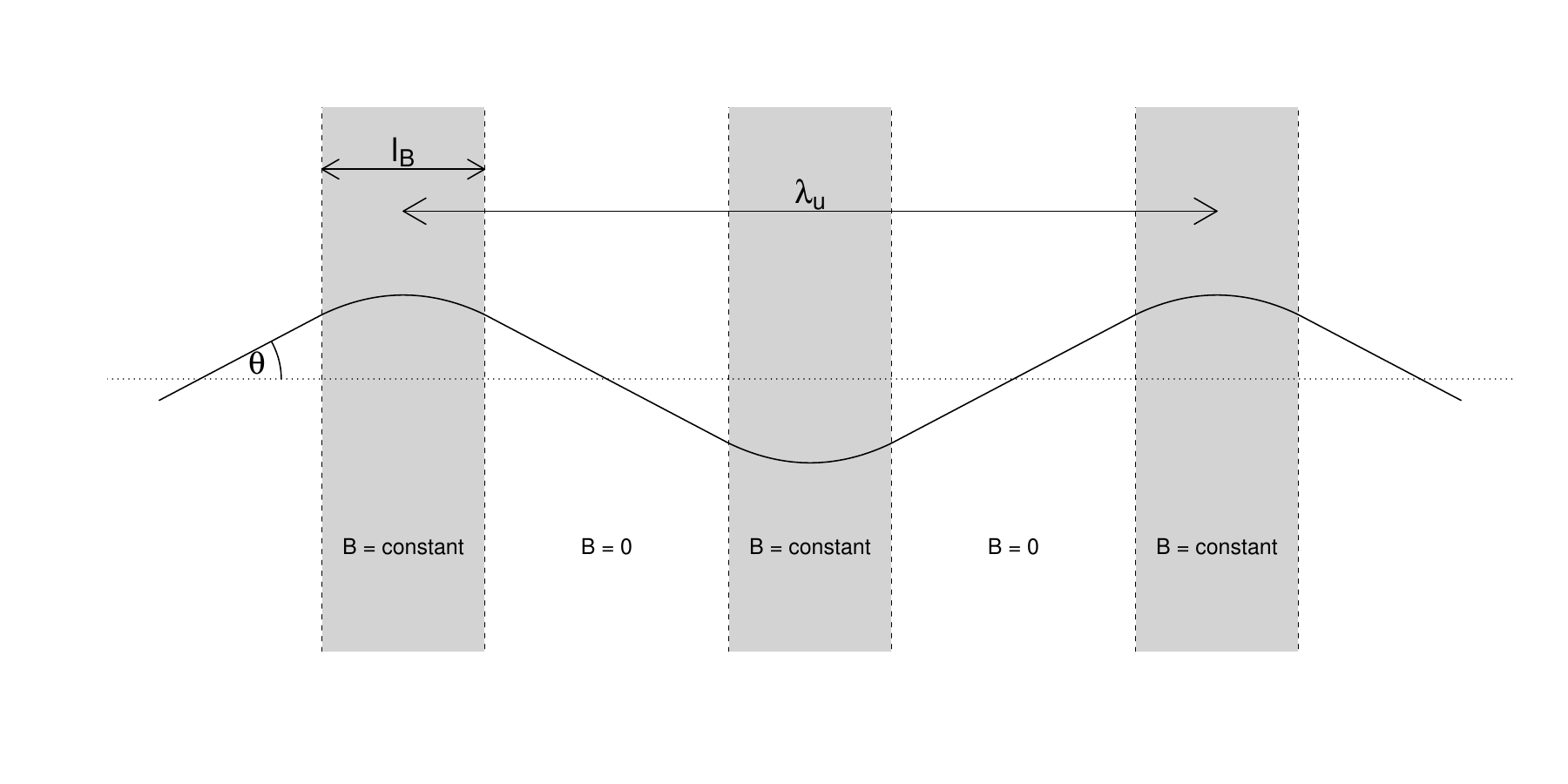}}
\caption{Possible undulator design.}
\label{undulator}
\end{figure}

To achieve a narrow bandwidth the number of periods can be increased by using a design of the type shown in Figure~\ref{undulator}, where the magnetic field is only non-zero for a small part of each undulator period. However, if $l_B$ is too small, pair production will not be suppressed. This is because the phase per undulator period, $\phi$, goes as $\theta^2 \propto l_B^2$ while the number of undulator periods goes as $N \propto \frac{1}{l_B}$. Thus for sufficiently small $l_B$ the total phase over the length of the undulator, $N\phi$, will become small and there will be no cancellation. This only occurs for very small $l_B$, however: pair production will be supressed as long as $N\phi \gtrsim 2\pi$, and for symmetrical electron-positron pairs\footnote{For \emph{asymmetric} pairs the two particles will have different phases and the slower particle can even have a phase that is a multiple of $2 \pi$; however, the faster particle will always have a phase that is between 0.5 and 1 times the phase for a symmetric pair and so the supression will still occur.} this is still satisfied for $l_B \approx 0.2$\,m, corresponding to $N \approx 16000$ and $\phi \approx 0.005$. For these parameters the bandwidth of the undulator would be $\sim \frac{1}{16000}$.

Nevertheless there will still be some pair-production, and we do not want the electrons and positrons to enter the interaction region. The same goes for any remaining muons. These particles can be swept aside by a short section of magnetic field. Since the bending angle is proportional to $B$ but pair production is proprotional to $B^{2/3}$ for strong fields~\cite{Uggerhoj} this field should be as intense as possible to prevent further pair production occuring within it. For example, a field of $10^4$ T and length 1 mm would bend the particles through an angle~$\approx 5 \times 10^{-15}$ with pair production losses of less than 0.2\%. The distance between the synchrotron conversion region and the interaction region is limited by the requirement that the photon beam does not become too wide. 
For the photon beam divergence discussed below, and assuming a maximum permissible spread of $10^{-9}$\,m, the distance to the interaction region can be up to a few times $10^9$\,m. The electrons, positrons and muons would thus be deflected by~$\sim 10^{-5}$\,m, which should be sufficient.

Apart from a narrow bandwidth, it is also important for the photon beam to have a small divergence. As we will see in section~\ref{interaction region}, this reduces the probability that secondary particles produced in $\gamma \gamma$ collisions will interact with the beams. The required divergence is very small, of order $3\times 10^{-19}$. Undulator radiation is indeed tightly beamed; for a single charged particle the photons are emitted into a cone of angular width approximately $1/\gamma$. However this is about $2 \times 10^{-16}$, so still much too large. 

The divergence at the resonant wavelength $\lambda$ does decrease as $1/{\gamma \sqrt{N}}$, so it appears having a large number of undulator periods would help here. However, in fact there is significant emission off-axis at wavelengths near
\begin{equation}
\lambda \approx \frac{\lambda_u}{2 \gamma^2} (1+K^2+\gamma^2\theta^2),
\label{Undulator off axis}
\end{equation} 
where K is the ``undulator parameter'' and $K \le 1$ for the parameters considered here. This gives a quadratic realtionship between the wavelength and the angle relative to the beam axis. For small $K$, significant emission occurs up to angles $1/\gamma$, corresponding to wavelengths up to twice the resonant wavelength in eq.~(\ref{Undulator equation}). Increasing the number of undulator periods sharpens the emission into a narrrow band around eq.~(\ref{Undulator off axis}) in wavelength-angle space but does not change the total bandwidth or divergence. The significant emission at angles up to $1/\gamma$ also means that the discussion above of how to achieve a narrow bandwidth while suppressing pair-production is not valid, since it assumed the radiation was on axis.

Current ``diffraction-limited'' light sources~\cite{Eriksson:vv0002} can in fact have much smaller divergences than this; the terminology ``diffraction-limited'' means that they can achieve divergences of order $\sigma_x/\lambda$, where $\sigma_x$ is the electron beam width. This can be much smaller than $1/\gamma$. Such small divergences can be achieved because the electrons radiate coherently, and constructive interference occurs only in the forward direction. Off-axis there is destructive interference and so the radiation is suppressed. A minimum requirement for this to work is that the radiation emitted by separate electrons must overlap in space. Since each electron emits a wavetrain of length $N$ periods as it passes through the undulator, the electron density must be high enough that there are multiple electrons within $N$ wavelengths of each other longitudinally.

This requirement is easily satisifed for modern light sources. Typically there are $\sim 10^{10}$ electrons in a bunch of length ~$\sim 0.01$\,m, the wavelength is $\sim 10^{-10}$\,m and $N \sim 100$, which means $\sim 10^4$ wavetrains overlap. However in our case even for N=16000 we would need a muon density over $\sim 10^{-25}\, \mathrm{m}^{-1}$, which seems extremely challenging.

To see how a divergence smaller than $\frac{1}{\gamma}$ might still be achieved, consider the narrow bandwidth (on-axis) for the radiation produced by a single particle passing through an undulator with many periods. This is produced by interference between the radiation emitted at different points in the undulator by the \emph{same} particle, not multiple particles. This can be understood as an example of the uncertainty principle: a photon detected far from the undulator could have been emitted anywhere along the undulator so $\delta z$ is large. Therefore the uncertainty on its momentum $\delta_{p_z}$, and hence on the wavelength, is small. Similarly, the divergence is proportional to the uncertainty on the transverse momentum $\delta_{p_x}$, so to achieve a small divergence the uncertainty on the transverse location where the photon was emitted, $\delta_x$, should be large.

For a diffraction-limited light source this uncertainty comes from the uncertainty over which particle in a distribution of width $\sigma_x$ emitted the photon, with the position of each particle being treated classically. For the single particle case it can instead come from \emph{quantum} uncertainty on the location of the particle. This concept has specifically been shown to increase the directionality of undulator radiation when using two interfering beams (i.e. splitting the spatial wavefunction into two)~\cite{Wong}.  

\begin{figure}[h]
\centerline{\includegraphics[width=0.6\linewidth]{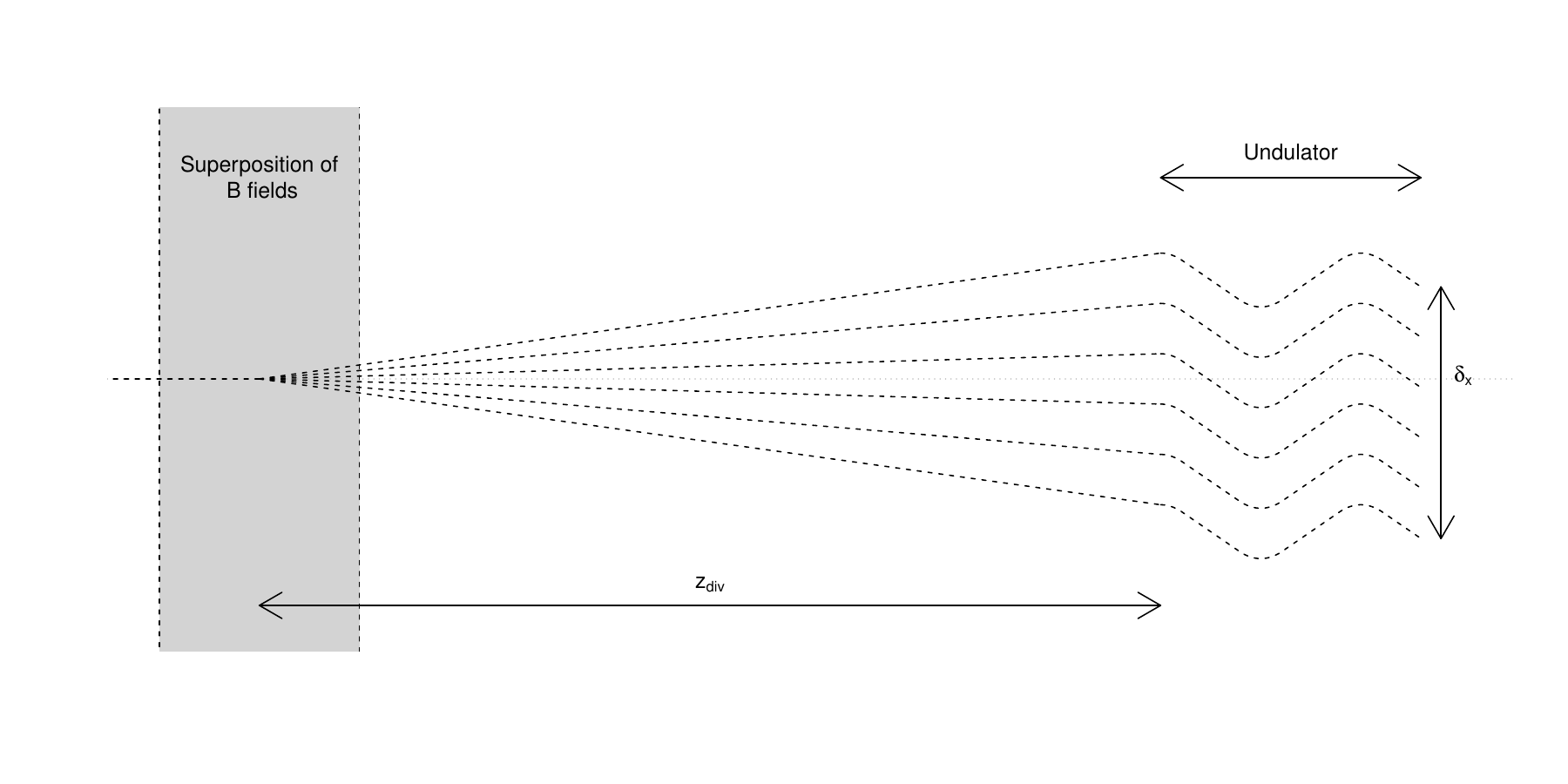}}
\caption{Arrangement of phased array and undulator. The muon trajectories could be recombined through a second region of superposed magnetic fields after the undulator (not shown) to avoid which-way information that would collapse the superposition.}
\label{phased array}
\end{figure}

If the wavepacket is split into only two components, there will still be substantial radiation into the side lobes of the interference pattern. To avoid this, the wavepacket should be split into a large number of components $N_x$. If these are spaced by $\delta_x/N_x$, the side lobes will occur at angles $\sim \frac{N_x \lambda}{\delta_x}$, and if this is greater than the inherent width of undulator radiation $1/\gamma$ there will be little power lost to the side lobes. We therefore need $N_x \gtrsim 700$. This can be achieved by splitting the wavefunction into two 10 or more times (since $2^{10} > 700$), with each split being half the size of the preceding one. The splitting is required in both the $x$ and $y$ directions, so a total of $\approx 20$ pairwise splits are required. The final result will be a square grid wavefunction which will essentially radiate as a phased array. The overall arrangement is shown in Figure~\ref{phased array}.

A variety of methods for splitting low-energy electron beams are summarised in~\cite{Wong}. It may be possible to apply one of these methods directly to the muon beam after it has been accelerated. If not, one option would be to split the wavefunction at low energies. The wavefunction would then be propagated through the accelerating stages of the collider~\cite{Brooks:2018vye}. If this is also not possible (e.g. due to decoherence during acceleration), the required superpositions could be produced by passing the beam through a superposition of magnetic fields produced by a SQUID. This device can superpose two states which differ by a single quantum of magnetic flux, $\Phi \approx 2.1\times 10^{-15}\,\mathrm{ T m^2}$. The corresponding magentic fields, although very small, would suffice as long as the distance between the point the trajectories diverge and the undulator, $z_\mathrm{div}$, is large enough, and we have plenty of space. The advantage of using SQUIDs is that they can be positioned slightly away from the beam and so will perturb the muons less than some of the other methods.

Note that if a very low divergence can indeed be achieved the bandwidth can also be made small by increasing the number of undualtor periods, since the assumption that the radition is on-axis will be correct.

In conclusion, the most feasible design for a GUT-scale collider is probably a photon-photon collider, with the photons produced by synchrotron radiation from a muon beam of slightly higher energy than the desired photon energy. By using a suitably designed undulator, and splitting the muon wavefunctions in the transverse direction, a narrow bandwith, small angular width, and a low pair production rate can be achieved.
\end{multicols}

\begin{multicols}{2}

\section{Interaction region}
\label{interaction region}
The low photon-photon cross-section, which we have gone to all this effort to achieve, now becomes a problem: the probability for any two photons in the colliding beams to interact will be small. Since photons are neutral, they cannot be focused to another collision point, so they must have a high probability of interacting in a single collision otherwise the energy used producing them will be wasted. Therefore, to make the interaction probability large, bunches with large numbers of particles will be required. This means many secondary particles can be produced, and can potentially interact with other photons and with each other. It turns out this can lead to serious problems.

The key reactions, and their cross-sections at $\sqrt{s} = 10^{15}$\,GeV, are ($h$ here stands for a hadron):
\begin{itemize}
\item{$\gamma \gamma \to \mathrm{hadrons}$, $12\, \mathrm{\mu b}$~\cite{Agashe:2014kda}.}
\item{$\gamma \gamma \to e^+e^-e^+e^-$, $6\, \mathrm{\mu b}$~\cite{Ginzburg:1981vm}.}
\item{$\gamma e \to ee^+e^-$, 150\,mb~\cite{Ginzburg:1981vm}.}
\item{$\gamma h \to \mathrm{hadrons}$, 4\,mb.}
\item{$h h \to \mathrm{hadrons}$, 1200\,mb.}
\label{Key reactions}
\end{itemize}

For the last two reactions the exact value of the cross-section will depend on the identity of the initial-state hadron(s). Here, to be specific, we have used the $pp$ cross-section in the last reaction, and used the fact that the ratios $\frac{\sigma_{\gamma h \to \mathrm{hadrons}}}{\sigma_{h h \to \mathrm{hadrons}}}$ and $\frac{\sigma_{\gamma \gamma \to \mathrm{hadrons}}}{\sigma_{\gamma h \to \mathrm{hadrons}}}$ are similar, and essentially measure the hadronic component of the photon (see e.g.~\cite{Godbole:2003yz}), to interpolate the cross-section for $\gamma h \to \mathrm{hadrons}$. We will also need to know how many hadrons are produced per collision in the first and last two processes. This is difficult to estimate, but around $10^4$ seems a plausible number. 

Now let us consider what will happen when two photon bunches collide. Initially the bunches consist only of photons and only the first two reactions are relevant. These will rapidly increase the proportions of electrons, positrons and hadrons: when the number of photons has fallen to $1-\epsilon$ of its inital value, the proportion of electrons and positrons will each be $2\epsilon / 3$, and the proportion of hadrons will be $10^4 \epsilon$. As the number of hadrons increases, the reaction $\gamma h \to \mathrm{hadrons}$ will start to become important; this will happen when the hadron proportion is~$\approx \frac{18}{4000}$, which is around $\epsilon \approx 5\times 10^{-7}$. Similarly, the reaction $\gamma e \to ee^+e^-$ will start to become important when the electron and positron proportions reach $\approx \frac{18}{300000}$, which is around $\epsilon \approx 10^{-4}$.

This is a disaster! Once these other reactions become important, they will produce yet more electrons, positrons and hadrons, all of which can react with the remaining photons. Effectively, the cross-section for the process $\gamma \gamma \to \mathrm{anything}$ is far higher than we expected. This in turn means that the energy requirements for a $\gamma \gamma$ collider are much higher.

However, this analysis is too pessimistic since it neglects the fact that the secondary particles are produced at an angle $\theta$ to the beams. They will thus remain in the collision region for only a fraction of its length. This is the reason alluded to above for the photon beams to be as narrow as possible. If photon beams of divergence $\theta_\mathrm{div}$ are collided head-on, the secondary particles will only travel through a fraction $f \sim \frac{\theta_\mathrm{div}}{\theta}$ of the collision region before exiting its side. This will reduce the probability for a secondary particle to collide with a beam photon by a factor $f$, and the probability for two secondary particles to collide by a factor $f^2$.

Furthermore, the effective cross-section can be further reduced by using the fact that these reactions do not take place at a single space-time point but over an extended region. Consider as an example the reaction $\gamma \gamma \to \mathrm{hadrons}$. At the parton level, this begins with each photon splitting into a $q\bar{q}$ pair, and then typically into multiple partons~\cite{Engel:1995yda}. Due to time dilation, this ``partonisation" takes place over a finite length scale $\sim \frac{E_\gamma}{\Lambda_{QCD}} \frac{1}{\Lambda_{QCD} } \sim 0.1$\,m. In a typical soft interaction the partonised photons then exchange pomerons and further partons in a volume of lengthscale $\frac{1}{\Lambda_{QCD}}$. Each colour-singlet combination of partons will produce a chain which will then fragment into hadrons. In the Lund string model~\cite{Andersson:1983ia} this hadronisation begins at the centre of the chain producing hadrons almost at rest within a timescale~$\frac{1}{\Lambda_{QCD}}$, and proceeds towards the ends of the chain producing hadrons with larger Lorentz boosts $\gamma$ after a timescale~$\sim \gamma \frac{1}{\Lambda_{QCD}}$ until it reaches the end of the chain where the fastest hadrons are produced in a timescale $\lesssim \frac{E_\gamma}{\Lambda_{QCD}} \frac{1}{\Lambda_{QCD}}$.

To exploit this behaviour to reduce the cross-section, we can impose a periodic magnetic field during the first stage of the collision. This will suppress the amplitude for partonisation, similarly to how we supressed the amplitude for pair production in the undulator. In this case there is a complication in that particles with different charges are involved: these will follow different trajectories in the magnetic field and hence accumulate different phases after one period of the field. Specifically, a particle of charge $Z|e|$, where $e$ is the electron charge, will be bent through an angle proportional to $Z$ in a magnetic field; thus the phase
\begin{equation}
\frac{2\pi}{\lambda}\int dl \frac{\theta^2}{2}
\label{phaseinbfield}
\end{equation}
is proportional to $Z^2$.
The pair-produced particles have charges $|Z| = \{1/3, 2/3, 1\}$ for down quarks, up quarks and electrons respectively; thus after one period of the magnetic field, their phases will be in the ratio 1~:~4~:~9.

We have investigated the reduction in amplitude that can be achieved in a toy model. We take the length of the magnetic field $l_B$ equal to the length of the partonisation region. The magnetic field has $N$ periods and is piecewise constant:
\begin{equation}
B = 
\begin{cases}
0, & x \le 0 \\
0, & x \ge l_B \\
c, & 0 < x < l_B \text{ and } |x-\frac{l_B}{N}| < \frac{l_B}{4} \\
-c, & 0 < x < l_B \text{ and } |x-\frac{l_B}{N}| \ge \frac{l_B}{4}
\end{cases}
\end{equation}
For a particle pair produced at $x$ we calculate the trajectory in the magnetic field from $x$ to the end of the magnet, and then use equation~(\ref{phaseinbfield}) to calculate the corresponding phase $\phi_x$. Finally we integrate over $x$:
\begin{equation}
\text{Production rate} \propto \left\lvert \int dx e^{i \phi_{x1} \phi_{x2}} \right\rvert ^2.
\label{productionrateequation}
\end{equation}
Here $\phi_{x1}$ and $\phi_{x2}$ are the phases for the two particles in the pair, which are in general different since the two particles can have different momenta. However, production of pairs with large invariant mass, and thus very different momenta, is suppressed~\cite{Ginzburg:1981vm}. We thus vary the momenta over the range $(p_1,p_2)=(0.1,0.9)p$ to $(p_1,p_2)=(0.9,0.1)p$. We weight all momenta within this range equally; in reality the amplitude will peak for equal momenta, but the importance of different momenta will also depend on the angle the resulting hadrons will be produced at, their energies etc., and for the purposes of this toy model we neglect all that.

\begin{figure}[h]
\centerline{\includegraphics[width=0.6\linewidth]{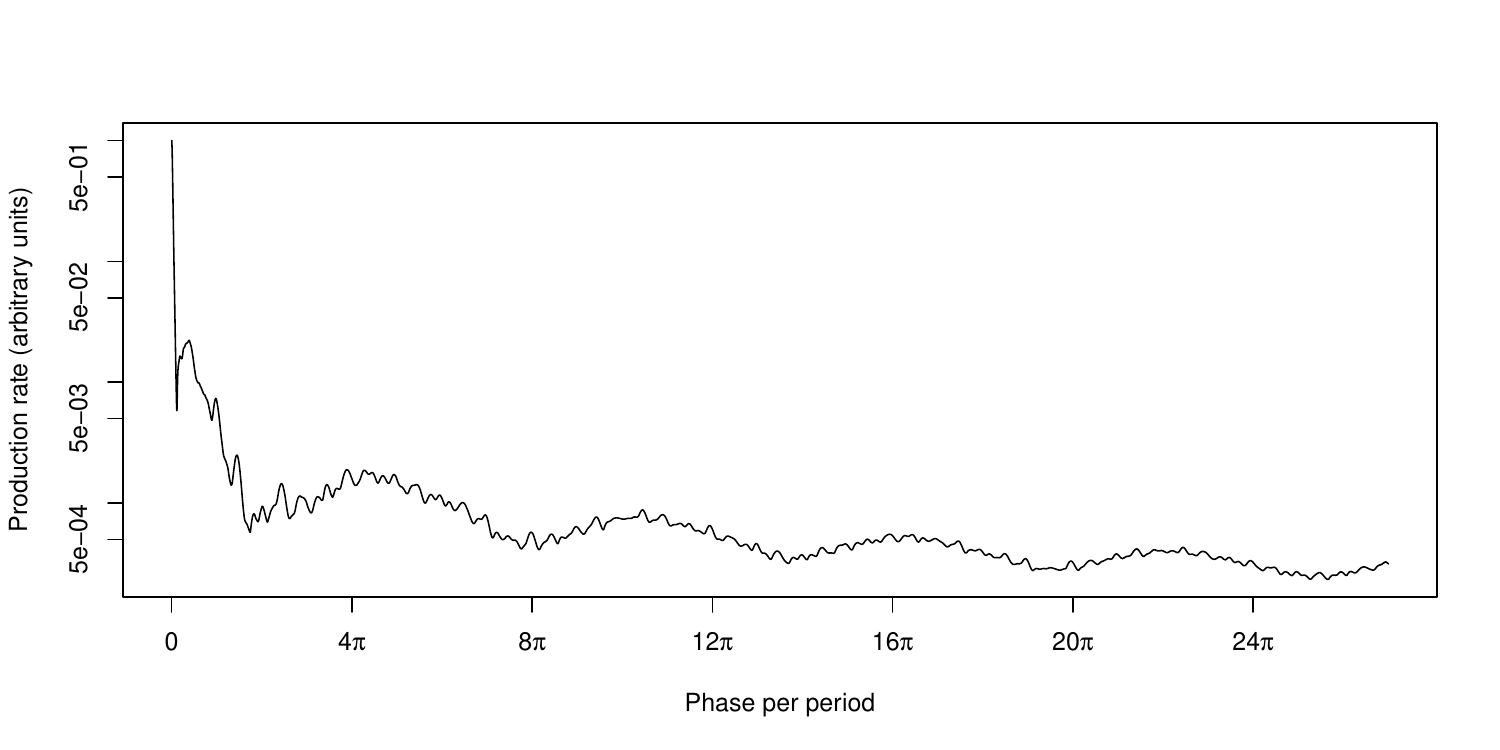}}
\caption{Suppression of pair production for $N=3$.}
\label{productionrate}
\end{figure}

We show the result for the case $N=3$ in Figure~\ref{productionrate}, where we parametrise the magnetic field by the phase a symmetric pair acquires over a single period of the field. It is clear that substantial suppression, to a level~$\sim0.0002$, is possible. However, as described above, the phases for down quarks, up quarks and electrons are different so this suppression cannot be achieved for all three simultaneously. If we wish to minimise the sum of the three suppression rates the best set of phases is $\{2.14, 8.56, 19.25 \} \times \pi$, where the sum of production rates is suppressed by a factor of 0.00049. For $l_B = 0.1$ m this corresponds to a magnetic field strength of 8900\,T.

\end{multicols}

\begin{multicols}{2}

This is a very strong magnetic field, but it is only required over a very small volume 0.1 m long and perhaps $10^{-9}$ m wide for less than a nanosecond. Fields of this strength (albeit longitudinal and constant in space rather than transverse and periodic) have already been produced by the Z Pulsed Power Facility over such volumes and timescales~\cite{Gomez2014}, so this should be feasible. They could also be produced by relatively low-energy electron bunches propagating alongside the photon beams; a 8900\,T magnetic field could be produced by a current of 4500\,A at a distance of $10^{-7}$\,m. Again, this is only required for less than a nanosecond, so the electron bunch needs to include $\sim 10^{13}$ electrons, which seems feasible.

A realistic analysis would have to include several complications. Firstly, the above calculation is for production of a single pair, but the first two reactions produce two pairs, so there will be additional suppression. We have also implicitly assumed suppression of each species is equally important, but in fact the production rate for down quarks is lower than for up quarks by a factor of 16.
Electron-positron pair production has a very long formation length $\gtrsim 10^5$ m at these energies; thus the reactions involving electrons and positrons could be further suppressed by extending the magnetic field over a longer lengthscale.

Another point is that we have carried out the above calculations for a fixed photon momentum $5 \times 10^{14}$ GeV. This is a good approximation for the first four reactions. However the last reaction will occur at a range of lower energies, and involve a hadron partonising\footnote{Partonisation is not required for valence quarks, which are there all along, but these only contribute a small part of the cross-section.}, rather than a photon. In this case the supression will be much weaker. The partonisation lengthscale is much shorter, say $10^{-3}$\,m for hadrons with energy $5 \times 10^{12}$ GeV, and in particular is much shorter than a single period of the magnetic field. Thus the hadron will experience a constant magnetic field. The partonisation rate will be suppressed by approximately the ratio of the partonisation lengthscale and the distance over which quarks are bent through an angle $\frac{\Lambda_{QCD}}{E} \sim 10^{-13}$, which is about $10^{-4}$\,m. So the partonisation rate will be suppressed by roughly a factor of 10.

Finally we have neglected higher order reactions and those which produce heavier particles. These have smaller reaction rates but also much shorter formation lengths which mean they will be suppressed much less.

Some of these complications will increase the suppression and some will decrease it. In the absence of a full calculation we will assume that suppression by about a factor of 300 is possible for the reactions $\gamma \gamma \to \mathrm{hadrons}$ and $\gamma h \to \mathrm{hadrons}$, that the reactions involving electrons can be neglected, and that the cross-section for secondary hadrons to interact, $h h \to \mathrm{hadrons}$, is reduced by about a factor of 10.

\subsubsection*{Maximum divergence}

In order to examine how the cascade of secondary and subsequent reactions depends on the divergence we also need to know how many particles will be produced in each collision, and what angles they will be produced at. At extremely high energies, colliding photons behave very much as hadrons with a reduced cross-section~\cite{Engel:1995yda}. In particular, this means that most collisions are ``soft" -- that is, the momentum and energy exchanged are small -- and the products emerge predominantly along the beam directions. To be specific, for proton-proton collisions, an approximately flat distribution in \emph{rapidity} is expected. This is almost the same as a flat distribution in \emph{pseudorapidity}, $\eta = -\mathrm{ln} (\mathrm{tan} \frac{\theta}{2})$, where $\theta$ is the angle to the beam direction~\cite{Schegelsky:2010xi}. I have not been able to find a clear statement in the literature on the expectations for photon-photon collisions, but simulations show similar behaviour~\cite{Badelek:2001xb}. Also, similarly to the case for hadrons, the simulations in~\cite{Badelek:2001xb} show that most events have very low transverse momenta, of the order of 1 GeV, so for large $\eta$ (small $\theta$), $\theta \approx \frac{1 \mathrm{GeV}}{E}$, where $E$ is the particle energy.

The distribution in $\eta$ is approximately flat up to the maximum kinematically allowed pseudorapidity $\approx \mathrm{ln}\frac{\sqrt{s}}{\Lambda_{QCD}}$, which we take to be 35 for the reactions $\gamma \gamma \to \mathrm{hadrons}$ and $\gamma h \to \mathrm{hadrons}$ and 27.5 for $h h \to \mathrm{hadrons}$. We take the number of hadrons per unit pseudorapidity to be 100 in all cases. The most important hadrons are those produced at large pseudorapidities, since these have the most energy and also will stay in the beam the longest. We must also consider the effect of the magnetic field. This bends the partons through an angle $\theta_\mathrm{bend}$ which depends on the location where partonisation began, but is of order $5\times10^{-14}$ for the partonisation of primary photons, and $10^{-12}$ for the partonisation of hadrons with energy $\sim10^{12}$\,GeV. The hadrons eventually produced by these partons will inherit this angle\footnote{This is true for neutral hadrons. Charged hadrons will be bent through a further angle; we neglect this.}. So hadrons with the largest pseudorapidities, above $-\mathrm{ln} (\mathrm{tan} \frac{\theta_\mathrm{bend}}{2})$, will end up travelling at angles $\sim \theta_\mathrm{bend}$ rather than the angles corresponding to their pseudorapidities. This occurs for pseudorapidities above about 30 for $\gamma \gamma \to \mathrm{hadrons}$ and $\gamma h \to \mathrm{hadrons}$ and 27.5 for $h h \to \mathrm{hadrons}$. Thus around 500 and 100 hadrons respectively will be produced with angles around $\theta_\mathrm{bend}$. We neglect hadrons produced at larger angles.

In summary, the reactions in the magnetic field have approximately the following properties:
\begin{itemize}
\item{$\gamma \gamma \to \mathrm{hadrons}$: $\sigma = 40\,\mathrm{nb}$, producing $\sim 500$ hadrons at $\theta \approx 5\times10^{-14}$.}
\item{$\gamma h \to \mathrm{hadrons}$: $\sigma = 13\,\mathrm{\mu b}$, producing $\sim 500$ hadrons at $\theta \approx 5\times10^{-14}$.}
\item{$h h \to \mathrm{hadrons}$: $\sigma = 100\,\mathrm{mb}$, producing $\sim 100$ hadrons at $\theta \approx 10^{-12}$.}
\label{Suppressed reactions}
\end{itemize}

We can now calculate the divergence required to avoid a cascade of secondary and further reactions. If the number of photons per unit area is $n_\gamma$, the probability of a photon interacting with another photon is $p_{\gamma \gamma} = \sigma_{\gamma \gamma \to \mathrm{hadrons}} n_\gamma$, and the number of secondary hadrons per unit area will be
\begin{equation}
n_h \approx 500 \sigma_{\gamma \gamma \to \mathrm{hadrons}}  \frac{\theta_\mathrm{div}}{5\times10^{-14}} n_\gamma^2.
\label{secondaryhadronsequation}
\end{equation}
The probability of a photon to interact with one of these secondary hadrons is then $p_{\gamma h} = \sigma_{\gamma h \to \mathrm{hadrons}} n_h$. We require $p_{\gamma h} < p_{\gamma \gamma}$ to avoid a cascade, and also $p_{\gamma \gamma} \approx 1$ so that most photons interact; this gives the requirement $\theta_{\mathrm{div}} < 3\times 10^{-19}$ and also $n_\gamma \approx 1/\sigma_{\gamma \gamma \to \mathrm{hadrons}}$ which will set the photon density. It is straightforward to check that if these requirements are satisfied the probability for two secondary hadrons to interact is small and the number of tertiary hadrons produced in such collisions is also small. 

This calculation shows that the exponential cascade can be entirely avoided and so the effective cross-section is just equal to the photon-photon cross-section in the periodic magnetic field, $\sigma_{\gamma \gamma \to \mathrm{hadrons}}$. For the remainder of this paper we will assume that a cross-section of around $40\,\mathrm{nb}$ can in fact be achieved. Note that this a factor $3 \times 10^7$ lower than the \emph{pp} cross-section we started with.

With this cross-section, the energy required to collect 10 interesting events is about 1.5 years of the total solar luminosity. Since the collider will likely take decades to construct (see section~\ref{location}) it would probably make sense to run for longer, say 30 years, using a fraction, say 5\%, of the solar luminosity.

\subsubsection*{Neutrino collisions}
We can now return to our claim above that the larger divergence for (anti-)neutrino beams produced at the $Z^0$ pole will cause unavoidable problems. Recall that in that case the divergence will be roughly $10^{-13}$. The difference to the case of photons is that the divergence cannot be reduced by splitting the $Z^0$ wavefunction spatially. 

The reason for this is somewhat subtle. Essentially, it is because the $Z^0$ decay produces a neutrino-antineutrino \emph{pair} as opposed to the single photon produced in an undulator. When summing over the components of the $Z^0$ wavefunction spaced by a distance $d$ in $x$, at large distances there will be a phase linear in $x$ for a neutrino observed at an angle $\theta$. For angles not satisying Bragg's law $\sin \theta = \frac{n \lambda}{2d}$ there would be destructive intereference when these phases are summed; however, this fails to take into account the antineutrino, which by momentum conservation is produced at an angle $-\theta$, producing an exactly opposite phase. Therefore the total phase for the pair is independent of $\theta$ and there is no destructive intereference and hence no reduction in the divergence. 

To assess the impact of this divergence we also need to know the cross-sections for interactions involving neutrinos\footnote{After filtering out quarks and leptons, the $Z_0$ decay will produce a beam containing equal numbers of neutrinos and antineutrinos. Therefore $\nu\nu$, $\nu\bar{\nu}$ and $\bar{\nu}\bar{\nu}$ interactions are all relevant, but for simplicity we shall just refer to "neutrinos" in the remainder of this section.}, the number of particles produced and the angles they are produced at.

The process with the highest cross-section at very high energies is $\nu\bar{\nu} \to l\bar{l}$, for which the cross-section is 220 pb~\cite{Lacki:2015lda}. However, as we have seen for photons, processes involving hadrons are more dangerous since they produce hundreds or more of hadrons at very small angles. As far as I am aware the cross-section for the process $\nu \nu \to \mathrm{hadrons}$ has not been calculated. To get a rough estimate, we will assume that it is dominated by a hadronic component of the neutrino, which is present because a neutrino can fluctuate to $l W$ and then to $l q\bar{q}$. This then implies $\frac{\sigma_{\nu \nu \to \mathrm{hadrons}}}{\sigma_{\nu h \to \mathrm{hadrons}}} \approx \frac{\sigma_{\nu h \to \mathrm{hadrons}}}{\sigma_{h h \to \mathrm{hadrons}}}$, in analogy to photons.

To estimate $\sigma_{\nu h \to \mathrm{hadrons}}$ we will use the neutrino-nucleon cross-section. This is conventionally calculated as a function of neutrino energy in the nucleon centre-of-mass frame; for a centre of mass energy $\sqrt{s}=10^{15}$\,GeV the neutrino energy $E_\nu$ is $5\times10^{29}$\,GeV. Different methods for calculating this cross-section start to diverge well below this, around $E_\nu = 10^{12}$\,GeV~\cite{Albacete:2015zra}. However, we will only require a lower limit. Examining Figure 5 of ref.~\cite{Albacete:2015zra}, a value of $1 \mathrm{\mu b}$ seems reasonable. Taking $\sigma_{h h \to \mathrm{hadrons}} \approx 1200 \mathrm{mb}$ as before, we then have $\sigma_{\nu \nu \to \mathrm{hadrons}} \approx 0.8 \mathrm{pb}$. 

For the first reactions $\nu \nu \to \mathrm{hadrons}$ and $\nu h \to \mathrm{hadrons}$, which proceed through $W^\pm$ and $Z^0$ boson exchange, the formation length is roughly $\frac{E}{m_V^2} \approx 10^{-4}$\,m. This is probably too short for the cross-section to be reduced in a periodic magnetic field as discussed above. As before we will assume that the cross section for $h h \to \mathrm{hadrons}$ can be reduced by about a factor of 10. In terms of numbers of hadrons produced, for the first two reactions all those produced at angles less than $\theta_{div}$ will remain in the beam for the whole length of the interaction region. Making the same assumptions as above the number of hadrons produced at such angles is about 500 in each case. The third reaction is as before. All this gives the following: 
\begin{itemize}
\item{$\nu \nu \to \mathrm{hadrons}$: $\sigma = 0.8\,\mathrm{pb}$, producing $\sim 500$ hadrons at $\theta \lesssim 10^{-13}$.}
\item{$\nu h \to \mathrm{hadrons}$: $\sigma = 1\,\mathrm{\mu b}$, producing $\sim 500$ hadrons at $\theta \lesssim 10^{-13}$.}
\item{$h h \to \mathrm{hadrons}$: $\sigma = 100\,\mathrm{mb}$, producing $\sim 100$ hadrons at $\theta \approx 10^{-12}$.}
\label{Neutrino reactions}
\end{itemize}

To avoid an exponential cascade, the hadron density must be low enough that each hadron produces one or fewer hadrons through the third reaction. A short calculation shows that this translates to $n_\nu \lesssim 10^{32}\,\mathrm{m}^{-2}$. This corresponds to an effective cross-section $\sigma_\mathrm{eff} =\frac{1}{n_\nu} \approx 100 \mathrm{\mu b}$ --- eight orders of magnitude larger than the cross-section for $\nu \nu \to \mathrm{hadrons}$, and three orders of magnitude larger than our estimate for the photon-photon cross-section in a periodic magnetic field. In fact, it is even above the unsuppressed photon-photon cross-section. So even if supressing the cross-section is not possible, a photon collider is still preferable to a neutrino collider.

\section{Detector design}
\label{detector}
As discussed at the beginning of section~\ref{interaction region}, because of the low photon-photon cross-section, bunches with large numbers of particles will be required. This in turn means that each bunch collision will release a lot of energy. 
For a beam width of $10^{-9}$\,m, the number of photons per bunch would have to be around $3\times10^{17}$, and the energy released per collision would be $5\times10^{22}$\,J. Designing the detector to absorb this energy will be a challenge.

As dicussed above, most of the collisions will be ``soft" and will have an approximately flat distribution in pseudorapidity up to the maximum kinematically allowed pseudorapidity $\approx \mathrm{ln}\frac{\sqrt{s}}{\Lambda_{QCD}} \approx 35$. This implies that the energy is distributed exponentially, $\frac{dE}{d\eta} \propto e^\eta$, with most of the energy concentrated in the interval $34 \lesssim \eta \lesssim 35$. Since these are soft events, they will not teach us anything about GUT-scale physics. Therefore one possibility would be to simply allow the collision products at high pseudorapidity to escape into space. However, if they can be absorbed, their energy can be recycled to improve the overall efficiency of the collider. This appears to be feasible for a carefully designed absorber.

When a high energy particle hits a detector, a shower of secondary particles is produced. The absorber must be large enough to contain this shower. The required absorber size depends on the type of particle: at moderate energies, electrons and photons require the smallest detector depths, followed by hadrons, and then muons. Neutrinos interact very weakly and it is impractical to contain them. However, at GUT-scale energies this changes: in ice at $\sim 10^{14}$\,GeV, shower depths are around 1 km for electrons and photons, tens of km for muons, but only $\approx 40$\,m for hadrons~\cite{Gerhardt:2010bj}. Neutrinos interact much more strongly at these energies and a depth of tens of km is also sufficient to absorb them~\cite{Albacete:2015zra}.

A relevant factor for the design of the detectors is the bunch structure. The photons in each beam can either be in a single long but relatively low-density bunch, or in a number of shorter denser sub-bunches which pass through each other, such that a photon will interact roughly once after passing through all of them. This second arrangement is probably more convenient: it means that collisions only occur in a number of discrete regions, so the low-$\eta$ detectors (which will be sensitive to hard collisions) are only needed in these regions and not along the whole interaction region. Also, the cross-section suppressing magnetic fields can be varied along the interaction region: within 0.1\,m of the collision regions they can be as described above, and further away they can be optimised to suppress only reactions involving electrons, which have a much longer formation length. Finally, splitting each bunch into a large number of sub-bunches reduces the energy that must be absorbed in one go in the detectors.

The number of sub-bunches must be large so that the secondary and tertiary hadrons pass through multiple sub-bunches and gaps before they exit the beam; otherwise they will see an above-average particle density and the effective cross-section will increase. Since the tertiary hadrons have $f\approx 3 \times10^{-7}$ this means there must at least around $3 \times 10^6$ sub-bunches within each bunch.

\subsection{Hadron absorber}
\label{hadronabsorber}
The most challenging part of the detector will be the part which absorbs hadrons, which carry most of the energy and have the shortest absorption length, at very high $\eta$, where most of the energy is concentrated. The absorber must not only absorb this energy but also convert it into a useful form capable of being transferred over large distances back to the accelerator. This is particularly challenging for hadronic absorption, since the small absorption depth means the energy density absorbed could be huge. To avoid this, the density of the absorber could be reduced to spread the absorption over a larger volume; however, it would then be difficult to extract heat from the interior of the absorber sufficiently fast. One way to deal with this is to separate the hadron absorber into two components: a small dense target which absorbs the beam energy and radiates it as X--rays, surrounded by a much larger shell which absorbs the X--rays and converts the energy into a useful form.

We begin by considering the dense target. This should be a cylinder about 40 interaction lengths in depth, and one interaction length in radius~\cite{AkchurinWigmans}. The nuclear interaction length depends on the density, 
$\lambda = k / \rho$, where $k \approx 100\,\mathrm{g}\,\mathrm{cm}^{-2}$ for most materials. Thus the number of atoms in the target scales as $\rho^{-2}$. Therefore at low temperatures, where most of the energy ends up in ions and electrons, the temperature goes as $\rho^2$ and the energy density in ions and electrons goes as $\rho T \propto T^{3/2}$. The energy density in photons increases faster; it goes as $T^4$. There is thus a crossover temperature, which is hundreds of eV for most light elements, above which most of the energy ends up in photons. In this regime the energy density in photons continues to increase as $T^4$ while the energy density in ions and electrons increases as $T^{7/3}$.

This means that if we choose the target parameters so that the temperature after the hadrons are absorbed is above the crossover temperature, most of the energy will indeed end up in X--ray photons. To obtain photons with a desired X--ray temperature, the volume of the absorber should be chosen so that the energy density of black-body radiation at that temperature, multiplied by the absorber volume, is equal to the absorbed energy. For example, suppose the beam width is $10^{-9}$\,m and there are $3\times10^6$ sub-bunches, so the energy to be absorbed per sub-bunch is about $10^{16}$\,J. If we require a temperature of 4 keV 
the volume of the absorber is about $2\,\mathrm{m}^3$. This in turn determines the interaction length and hence the density and mass of the absorber. The absorber turns out to be rather small, with a mass of $\sim 10^4$\,kg, a radius of tens of centimetres, and a length of ten metres or so.

We now turn to the large shell, and in particular the case where the energy is released as infra-red radiation; this has the advantage that it may be feasible to beam it directly to where it is required without requiring a further conversion step (e.g. a heat engine powering a laser). To minimise the sphere area it should be constructed of a material with a very high melting point such as graphite. The size of the sphere required seems manageable: for example, for a temperature of $\approx 4000$ K the radius would be $\approx 3000$ km. The thickness of the sphere is determined by the X--ray energy; it must be such that the absorption depth for the X--rays is smaller than the thickness of the graphite shell. For the particular case of graphite, a thickness of $100\, \mathrm{\mu m}$ is required for an X--ray energy of 4 keV~\cite{PDGproperties, HubbellSeltzer}. This corresponds to a mass $m \approx 3\times10^{13}$\,kg.

The total power used for collisions determines the total number of absorbers required: if 5\% of the solar luminosity is used, with 100\% efficiency (see also section~\ref{efficiency}), there would be $\approx 5000$ interaction points, and so $\approx 10^4$ absorbers, requiring a total carbon mass of around $3\times10^{17}$\,kg. This is modest: it is only about 0.1\% of the total carbon in the atmosphere of Venus.

\subsection{Other components}

It seems that a similar design to the hadronic absorber could be used to absorb electromagnetically interacting particles. Since their interaction length is longer at these energies, this should be placed behind the hadronic absorber. However, it is in fact difficult to use the same principle of a small dense target first absorbing the energy and then reradiating it as X--rays. This is because the EM absorber should be larger (because of the longer interaction length) but would absorb less energy, so the temperature it reaches would be too low for most of the energy to be in photons. A more elegant alternative is to place a weak magnetic field in front of the hadron absorber. This will cause a shower of electromagnetic particles down to energies for which $\frac{E}{m_ec^2}B \approx B_0$, similar to the ``preshower'' of ultra-high energy cosmic rays in Earth's magnetic field~\cite{Homola:2013sya}. We can choose $B$ so the shower stops at energies around $10^9$\,GeV, for which the hadronic and electromagnetic interaction lengths are similar~\cite{Gerhardt:2010bj}. The electromagnetically interacting particles can then be efficiently absorbed in the hadronic absorber.

This just leaves muons and neutrinos. These will only contain a small fraction of the energy, so it should be easier to design absorbers for them. On the other hand, precisely because they only contain a small fraction of the energy, it may not be worth the trouble of trying to recover energy from them and they may simply be allowed to escape. 

At lower $\eta$, the energy density is exponentially lower. Therefore the detector design becomes much less challenging and the detector can be designed with particle identification and tracking in mind, rather than energy recovery. The smaller energy density means this part of the detector can be much more compact.

\section{Accelerator design}
\label{accelerator}

Given the above, the requirements for the muon accelerator are:
\begin{itemize}
\item{A muon energy around $6 \times10^{14}$\,GeV.}
\item{A (sub-)bunch interval of $3 \times10^{-6}$ s.}
\item{$10^{11}$ muons per bunch.}
\end{itemize}

A circular accelerator with this energy would have huge synchrotron radiation losses unless its radius was huge, $\gtrsim 10^{30}$\,m (larger than the Hubble radius). A linear accelerator can be much shorter than this so would definitely be preferred. With current conventional accelerator technology, gradients of 100\,MV/m are possible. It seems plausible that this could be increased to, say, 1\,GV/m by the time the collider proposed here is built, which would give a length of $6 \times10^{14}$\,m (about a light month).

It is possible advanced accelerator technology could reach much higher gradients than this. Possibilities include acceleration in dielectric structures, plasma wakefield accelerators, or acceleration in crystal channels (see~\cite{Shiltsev:2012zzc} for a review). These might achieve gradients of 1\,TV/m, which would reduce the required length to less than $10^{12}$\,m -- roughly the radius of Jupiter's orbit.

Note that the total length of the collider would be slightly over double the length calculated above, since two muon accelerators are required, one in each direction. An additional allowance must be made for the undulators and for the detectors, but these will be orders of magnitude smaller.

The accelerator must also be designed to avoid excessive losses due to absorption of virtual photons~\cite{Casher:1997rr}. These can be estimated using the Weizs\"{a}cker-Williams approximation. In the accelerator frame, the Coulomb field of the muon becomes a short electromagnetic pulse. The energy of this pulse in an annulus between radii $R$ and $2R$ from the beam is $\gamma \alpha / 2R$, and the photon energy is $\omega \approx \gamma/R$. For, say, $R = 0.1$\,m we find the pulse has an energy of about 50~MeV and $\omega \approx 10$\,GeV. Absorption lengths at 10 GeV are tens of $\mathrm{g}\,\mathrm{cm}^{-2}$
(with high-$Z$ materials having the shortest absorption length), and a density of $\sim 1 \mathrm{g}\,\mathrm{cm}^{-3}$ seems reasonable to allow for gaps between components etc., which would give an absorption length of tens of centimetres. This would lead to losses of around 100 MeV/m, safely lower than the likely acceleration gradient. However, large densities of high-$Z$ materials very near ($\lesssim 1$\,cm) to the beam should be avoided. (For electrons this problem would be orders of magnitude worse; this is another reason to prefer muons to electrons as the primary particles.)

The losses can be further reduced by making the bunches very short, so that the Coulomb fields of the muons overlap. If there are $10^{11}$ muons per bunch and the bunches are $10^{-6}$\,m long the Coulomb fields will overlap within a few centimetres of the beam. At greater radii, there will be a single pulse rather than one for each muon, and the relevant photon frequencies will be around the bunch length. This is below the plasma frequency for most metals, so the photons will be reflected rather than absorbed, greatly reducing the losses.

Note that ref.~\cite{Casher:1997rr} assumes that the beampipe must be solid and very thick to contain the pressure of the electromagnetic fields which accelerate the high energy particles, and so concludes that the inner radius of the beampipe must be very large to prevent large losses by absorption of virtual photons. However, this is too pessimistic: it assumes the pressure is static and must be balanced by static forces in the pipe, whereas in fact it will only be present while the very short bunches are passing, so it can be balanced by inertial forces (the pipe can `stretch' while the bunch passes). In a plasma wakefield accelerator there is no beampipe anyway.

We have so far discussed the accelerator as if there is only one. However, this is not the case; there will need to be a large number of accelerators running in parallel. If there is one pair per interaction point there will be about $10^4$ of them.

The materials required to construct the accelerators will be substantial. It is hard to know how much mass would be required per unit length, since this depends so much on the acceleration technology. Assuming, say, 100\,kg\,m$^{-1}$, and 10000 accelerators each of length $6 \times 10^{14}$\,m, the total mass required would be $6\times 10^{20}$\,kg. For comparison, the mass of Mercury is $3 \times 10^{23}$\,kg, so this seems manageable.

One final remark about the length: the most important consideration in the design of the collider is the energy efficiency. It may be therefore that a lower acceleration gradient, and hence longer accelerators, will be preferred if it is more efficient.

\subsection{Location}
\label{location}

It would make sense for the interaction points to lie as near the solar system as possible, since this will decrease the total distance construction materials will have to be transported. However, the accelerators and detectors will need to lie almost on a straight line to reduce synchrotron losses. The problem with this is that an initially linear arrangement will not remain straight if the components are orbiting (except for the courageous option of the components moving directly towards/away from the Sun and hence aiming one of the beams directly at the inner solar system). One solution is to give the whole collider an initial transverse velocity away from the Sun. The outer parts would continue moving at approximately constant velocity, while the central parts would be slowed by the Sun's gravity, and would need to be accelerated to overcome this. The total delta-v required for this over the 30-year lifetime of the collider is approximately $\frac{4000}{d^2}\,\mathrm{km}\,\mathrm{s}^{-1}$, where $d$ is the distance from the Sun in astronomical units. This is just 10\,km\,s$^{-1}$ or less for distances around Uranus's orbit or greater.

Much larger delta-vs would probably be required to construct the accelerator in a reasonable timescale. For the length of $6 \times10^{14}$ m considered above, velocities in excess of 600 km s$^{-1}$ would be needed for the outermost accelerator components in order to get them to the required location within 30 years. However, these velocities are only required for the ends of the accelerators; smaller velocities are enough for the inner parts and the detectors.

Are these speeds energetically feasible? Suppose a total mass of $6\times 10^{20}$\,kg must be accelerated to an average speed of 500\,km\,s$^{-1}$; the total kinetic energy required is then about $10^{32}$\,J. This is ``only'' 0.01 years of the total solar luminosity, and hence very small compared to the energy which will be required to run the accelerator. So this is OK.

\section{Energy transfer}
\label{energy}

The energy required by the collider can be collected by a Dyson swarm~\cite{Dyson1960, StarMaker} -- a large number of solar power collectors that partially or completely enclose the Sun. The energy required to achieve the desired integrated luminosity is about $10^{34}$\,J; a Dyson swarm collecting 5\% of the Sun's light could collect this much energy in 30 years.

In addition, time would be required to construct the Dyson swarm. The speed of construction can increase exponentially with time, as the energy from the partially completed swarm is used to manufacture further collectors. One estimate is that constructing a complete swarm would take 30 years and that the mass required would be less than the mass of Mercury~\cite{ArmstrongSandberg}.

The energy collected by the Dyson swarm must then be transmitted to the accelerator, up to $6 \times 10^{14}$\,m away. Two methods proposed for transmitting power over long distances in space are microwaves (transmitted by antennae and received by rectennas) and light waves (transmitted by lasers and received either by photovoltaics or by some sort of heat engine). The minimum size for the transmitters and receivers is given by the diffraction limit. For electromagnetic waves of wavelength $\lambda$, the transmitter and receiver sizes $s_r$ and $s_t$ are related to the distance $L$ by $s_r s_t \gtrsim L \lambda$. Even for the largest distances considered, the required transmitter and receiver sizes are tens of km for lasers and up to thousands of km for microwaves.

However, a diffraction-limited receiver would be exposed to huge power densities. The power being transmitted is 5\% of the total solar luminosity; for a receiver tens (thousands) of km across, the power density is $\sim 10^{16}$ ($10^{12}$)\,W\,m$^{-2}$. This seems far too high, so either the receiver should be larger than the diffraction limit, or there should be many of them. Whether laser or microwave transmission should be used will depend on which has better energy efficiency, and to some extent also on the mass required. The mass required will be at most similar to the mass required for the Dyson swarm, since the total power absorbed is the same.

A possible approach which might help efficiency would be to use the same wavelength for both power transmission and particle acceleration. Conventional accelerator technology uses microwaves, and plasma acceleration uses lasers. Therefore it would in principle be possible to redirect the incoming electromagnetic radiation directly into the accelerator, which would avoid two conversion steps: microwave (laser) to electrical, and electrical to microwave (laser).

\subsection{Energy efficiency}
\label{efficiency}
It is very difficult to estimate the energy efficiency of possible future technologies without knowing the details. Optimistically, however, and bearing in mind that the Carnot efficiency for recovering waste heat could be very high since temperatures in the outer solar system are very low, we might assume something like the following: 
\begin{itemize}
\item{$a=80\%$ of the solar luminosity is delivered to the accelerator.}
\item{The accelerator efficiency is $b=60\%$.}
\item{$c=90\%$ of the collision energy is recovered and delivered back to the accelerator.}
\end{itemize}  
The total efficiency with these assumptions is $\frac{ba}{1-bc} = 104\%$. Thus an efficiency around 100\% appears possible.

\section{Conclusions}
\label{conclusions}
We have shown that a feasible design for a GUT--scale collider is a photon collider, with the photons produced as undulator radiation from muon beams of slightly higher energy. Such a collider would be approximately $10^{15}$\,m long, and would require 30 years to collect $\mathcal{O}(10)$ interesting events if powered by 5\% of the full solar luminosity. In order to control the energy requirements, secondary reactions in the collision region must be suppressed; we have shown that this can be achieved if the muon wavefunctions are split spatially to produce a very narrow photon beam and if partonisation takes place in a strong periodic magnetic field. 

There do not appear to be any fundamental reasons why a collider of the type described in this paper could not be built. However, there is the issue of vacuum stability to consider. We are assuming no physics beyond the Standard Model; however, the Standard Model vacuum is probably metastable~\cite{Bednyakov:2015sca}, raising the possibility that GUT-scale collisions could cause vacuum decay. Even if it can be shown that this would not occur within the Standard Model, there is the possibility of new physics changing this conclusion.

No cosmic rays have been observed with energies near the GUT scale. However, a greater understanding of the sources of cosmic rays might show that cosmic rays with these energies do exist and that they sometimes collide within or near these sources; if enough collisions have occurred in our past lightcone this would show that vacuum stability is not a problem. Similarly, greater understanding of cosmology may show that the early universe reached a temperature above $T_\mathrm{GUT}$, which would allow a similar conclusion to be reached. This issue should be resolved before construction begins.

\section*{Acknowledgements}
I am grateful to Stephen Brooks for many useful coonversations.

\end{multicols}


\begin{thebibliography}{9}

\bibitem{Agashe:2014kda}
  K.~A.~Olive {\it et al.} [Particle Data Group Collaboration],
  %``Review of Particle Physics,''
  Chin.\ Phys.\ C {\bf 38} (2014) 090001.

\bibitem{Budnev:1974de}
  V.~M.~Budnev, I.~F.~Ginzburg, G.~V.~Meledin and V.~G.~Serbo,
  %``The Two photon particle production mechanism. Physical problems. Applications. Equivalent photon approximation,''
  Phys.\ Rept.\  {\bf 15} (1975) 181.
  doi:10.1016/0370-1573(75)90009-5

\bibitem{Lacki:2015lda}
  B.~C.~Lacki,
  %``SETI at Planck Energy: When Particle Physicists Become Cosmic Engineers,''
  arXiv:1503.01509 [astro-ph.HE].

\bibitem{Neuffer:2015nna}
  D.~Neuffer, M.~Palmer, Y.~Alexahin, C.~Ankenbrandt and J.~P.~Delahaye,
  %``A muon collider as a Higgs factory,''
  arXiv:1502.02042 [physics.acc-ph].

\bibitem{Ginzburg:1981vm}
  I.~F.~Ginzburg, G.~L.~Kotkin, V.~G.~Serbo and V.~I.~Telnov,
  %``Colliding gamma e and gamma gamma Beams Based on the Single Pass Accelerators (of Vlepp Type),''
  Nucl.\ Instrum.\ Meth.\  {\bf 205} (1983) 47.
  doi:10.1016/0167-5087(83)90173-4

\bibitem{Asner:2001vh}
  D.~Asner {\it et al.},
  %``Higgs physics with a gamma gamma collider based on CLIC I,''
  Eur.\ Phys.\ J.\ C {\bf 28} (2003) 27
  doi:10.1140/epjc/s2002-01113-3
  [hep-ex/0111056].

\bibitem{Bogacz:2012fs}
  S.~A.~Bogacz, J.~Ellis, L.~Lusito, D.~Schulte, T.~Takahashi, M.~Velasco, M.~Zanetti and F.~Zimmermann,
  %``SAPPHiRE: a Small Gamma-Gamma Higgs Factory,''
  arXiv:1208.2827 [physics.acc-ph].

\bibitem{Uggerhoj}
U.~Uggerhoj,
``Ultrarelativistic Particles in Matter,''
Disseration, University of Aarhus, Denmark, 2011.

\bibitem{Homola:2013sya}
  P.~Homola, R.~Engel, A.~Pysz and H.~Wilczyński,
  %``Simulation of ultra-high energy photon propagation with PRESHOWER 2.0,''
  Comput.\ Phys.\ Commun.\  {\bf 184} (2013) 1468
  doi:10.1016/j.cpc.2013.01.015
  [arXiv:1302.6939 [astro-ph.IM]].

\bibitem{Maruyama:2015bga}
  T.~Maruyama, M.~K.~Cheoun, T.~Kajino, Y.~Kwon, G.~J.~Mathews and C.~Y.~Ryu,
  %``Quantum Field Theoretic Treatment of Pion Production via Proton Synchrotron Radiation in Strong Magnetic Fields: Effects of Landau Levels,''
  Phys.\ Rev.\ D {\bf 91} (2015) 123007
  doi:10.1103/PhysRevD.91.123007
  [arXiv:1503.05635 [astro-ph.HE]].

\bibitem{Bulyak:2015yua}
  E.~Bulyak and N.~Shul'ga,
  %``Electron spectra and coherence of radiation in undulators,''
  arXiv:1506.03255 [physics.acc-ph].

\bibitem{Eriksson:vv0002}
M.~Eriksson, J~ F.~van~der~Veen and C.~Quitmann,
%Diffraction-limited storage rings {--} a window to the science of tomorrow
Journal of Synchrotron Radiation {\bf 21} (2014), 837
doi: 10.1107/S1600577514019286

\bibitem{Wong}
L.~J.~Wong, N.~Rivera, N., C.~Murdia, {\it et al.}
%Control of quantum electrodynamical processes by shaping electron wavepackets
Nat Commun {\bf 12} (2021), 1700
doi: 10.1038/s41467-021-21367-1

\bibitem{Brooks:2018vye}
S.~Brooks,
%``Potential and Issues for Future Accelerators and Ultimate Colliders,''
doi:10.18429/JACoW-IPAC2018-TUXGBD1
[arXiv:1811.12479 [physics.acc-ph]].

\bibitem{Godbole:2003yz}
  R.~M.~Godbole, A.~Grau, G.~Pancheri and A.~De Roeck,
  %``Predictions for the gamma gamma total cross-section in the TeV region: An Update,''
  hep-ph/0303018.

\bibitem{Engel:1995yda}
  R.~Engel and J.~Ranft,
  %``Hadronic photon-photon interactions at high-energies,''
  Phys.\ Rev.\ D {\bf 54} (1996) 4244
  doi:10.1103/PhysRevD.54.4244
  [hep-ph/9509373].

\bibitem{Andersson:1983ia}
  B.~Andersson, G.~Gustafson, G.~Ingelman and T.~Sjostrand,
  %``Parton Fragmentation and String Dynamics,''
  Phys.\ Rept.\  {\bf 97} (1983) 31.
  doi:10.1016/0370-1573(83)90080-7

\bibitem{Gomez2014}
  M.~R.~Gomez \emph{et al.},
  Phys.\ Rev.\ Lett.\ {\bf 113} (2014) 155003
  doi:10.1103/PhysRevLett.113.155003

\bibitem{Schegelsky:2010xi}
  V.~A.~Schegelsky, M.~G.~Ryskin, A.~D.~Martin and V.~A.~Khoze,
  %``A note on rapidity distributions at the LHC,''
  arXiv:1010.2051 [hep-ph].

\bibitem{Badelek:2001xb}
  B.~Badelek {\it et al.} [ECFA/DESY Photon Collider Working Group Collaboration],
  %``TESLA: The Superconducting electron positron linear collider with an integrated X-ray laser laboratory. Technical design report. Part 6. Appendices. Chapter 1. Photon collider at TESLA,''
  Int.\ J.\ Mod.\ Phys.\ A {\bf 19} (2004) 5097
  doi:10.1142/S0217751X04020737
  [hep-ex/0108012].

\bibitem{Albacete:2015zra}
J.~L.~Albacete, J.~I.~Illana and A.~Soto-Ontoso,
%``Neutrino-nucleon cross section at ultrahigh energy and its astrophysical implications,''
Phys. Rev. D \textbf{92} (2015) no.1, 014027
doi:10.1103/PhysRevD.92.014027
[arXiv:1505.06583 [hep-ph]].

\bibitem{Gerhardt:2010bj}
  L.~Gerhardt and S.~R.~Klein,
  %``Electron and Photon Interactions in the Regime of Strong LPM Suppression,''
  Phys.\ Rev.\ D {\bf 82} (2010) 074017
  doi:10.1103/PhysRevD.82.074017
  [arXiv:1007.0039 [hep-ph]].

\bibitem{AkchurinWigmans}
N.~Akchurin and R.~Wigmans,
Nuclear Instruments and Methods in Physics Research {\bf A 666} (2012) 80.

\bibitem{PDGproperties}
Particle Data Group,
Atomic and Nuclear Properties of Materials.
http://pdg.lbl.gov/2015/AtomicNuclearProperties/

\bibitem{HubbellSeltzer}
J.~H.~Hubbell and S.~.M.~Seltzer,
NISTIR 5632
https://www.nist.gov/pml/x-ray-mass-attenuation-coefficients.

\bibitem{Shiltsev:2012zzc}
  V.~D.~Shiltsev,
  %``High energy particle colliders: past 20 years, next 20 years and beyond,''
  Phys.\ Usp.\  {\bf 55} (2012) 965
  doi:10.3367/UFNe.0182.201210d.1033
  [arXiv:1205.3087 [physics.acc-ph]].

\bibitem{Casher:1997rr}
  A.~Casher and S.~Nussinov,
  %``Is the Planck momentum attainable?,''
  hep-th/9709127.

\bibitem{Dyson1960}
F.~J.~Dyson,
Science {\bf 131 (3414)} (1960) 1667.

\bibitem{StarMaker}
O.~Stapledon,
``Star Maker,''
Methuen, 1937.

\bibitem{ArmstrongSandberg}
S.~Armstrong and A.~Sandberg,
Acta Astronautica {\bf 89} (2013) 1.
doi:10.1016/j.actaastro.2013.04.002

\bibitem{Bednyakov:2015sca}
  A.~V.~Bednyakov, B.~A.~Kniehl, A.~F.~Pikelner and O.~L.~Veretin,
  %``Stability of the Electroweak Vacuum: Gauge Independence and Advanced Precision,''
  Phys.\ Rev.\ Lett.\  {\bf 115} (2015) no.20,  201802
  doi:10.1103/PhysRevLett.115.201802
  [arXiv:1507.08833 [hep-ph]].

\end{thebibliography}
\end{document}